\def\Version{{5.1}}




\message{ Assuming 8.5" x 11" paper }    

\magnification=\magstep1	          

\raggedbottom

\parskip=9pt

%

\def\singlespace{\baselineskip=12pt}      
\def\sesquispace{\baselineskip=18pt}      









\font\openface=msbm10 at10pt
 %

\def\Reals         {{\hbox{\openface R}}}

 %
 %
 %







\def\Re  {\mathop{\rm Re}  \nolimits}    

\def\im  {\mathop{\rm im}  \nolimits}    




%

%

%
%



\def\implies{\Rightarrow}

%




\def\sqr#1#2{\vcenter{
  \hrule height.#2pt 
  \hbox{\vrule width.#2pt height#1pt 
        \kern#1pt 
        \vrule width.#2pt}
  \hrule height.#2pt}}


\def\dal{\mathop{\,\sqr{7}{5}\,}}
\def\block{\dal}



\def\lto{\mathop
        {\hbox{${\lower3.8pt\hbox{$<$}}\atop{\raise0.2pt\hbox{$\sim$}}$}}}
\def\gto{\mathop
        {\hbox{${\lower3.8pt\hbox{$>$}}\atop{\raise0.2pt\hbox{$\sim$}}$}}}
%
%
%


\def\half{{1 \over 2}}


\def\part{\subseteq}		




\def\to{\mathop\rightarrow}	

\def\orthog{\mathop\bot}


\def\SetOf#1#2{\left\{ #1  \,|\, #2 \right\} }



\def\interior #1 {  \buildrel\circ\over  #1}     



\def\grad{\nabla}


\def\basisvector#1#2#3{
 \lower6pt\hbox{
  ${\buildrel{\displaystyle #1}\over{\scriptscriptstyle(#2)}}$}^#3}

\def\alfa{\alpha}


\def\tilde{\widetilde}		
\def\bar{\overline}		







\fontdimen16\textfont2=2.5pt
\fontdimen17\textfont2=2.5pt
\fontdimen14\textfont2=4.5pt
\fontdimen13\textfont2=4.5pt 

\let\miguu=\footnote
\def\footnote#1#2{{$\,$\parindent=9pt\baselineskip=13pt%
\miguu{#1}{#2\vskip -7truept}}}
 %

\def\linebreak{\hfil\break}
\def\lbr{\linebreak}
\def\pagebreak{\vfil\break}


\def\BulletItem #1 {\item{$\bullet$}{#1 }}
\def\bulletitem #1 {\BulletItem{#1}}

\def\PrintVersionNumber{
 \vskip -1 true in \medskip 
 \rightline{version \Version} 
 \vskip 0.3 true in \bigskip \bigskip}

\def\author#1 {\medskip\centerline{\it #1}\bigskip}

\def\address#1{\centerline{\it #1}\smallskip}

\def\furtheraddress#1{\centerline{\it and}\smallskip\centerline{\it #1}\smallskip}

\def\email#1{\smallskip\centerline{\it address for email: #1}} 

\def\AbstractBegins
{
 \singlespace                                        
 \bigskip\leftskip=1.5truecm\rightskip=1.5truecm     
 \centerline{\bf Abstract}
 \smallskip
 \noindent	
 } 
\def\AbstractEnds
{
 \bigskip\leftskip=0truecm\rightskip=0truecm       
 }

\def\section #1 {\bigskip\noindent{\headingfont #1 }\par\nobreak\smallskip\noindent}

\def\subsection #1 {\medskip\noindent{\subheadfont #1 }\par\nobreak\smallskip\noindent}
 %

\def\ReferencesBegin
{
 \singlespace					   
 \vskip 0.5truein
 \centerline           {\bf References}
 \par\nobreak
 \medskip
 \noindent
 \parindent=2pt
 \parskip=6pt			
 }
 %


\def\reference{\hangindent=1pc\hangafter=1} 

\def\ref{\reference}

\def\sepref{\parskip=4pt \par \hangindent=1pc\hangafter=0}
 %

\def\journaldata#1#2#3#4{{\it #1\/}\phantom{--}{\bf #2$\,$:} $\!$#3 (#4)}
 %

\def\eprint#1{{\tt #1}}

\def\arxiv#1{\hbox{\tt http://arXiv.org/abs/#1}}
 %


\def\webhome{{\tt http://www.pitp.ca/personal/rsorkin/}}
 %

 %



\def\webtilde{\lower2pt\hbox{${\widetilde{\phantom{m}}}$}}

 %

\def\hpf#1{\webhome{\tt{some.papers/}}}
 %

\def\hpfll#1{\webhome{\tt{lisp.library/}}}
 %



\font\titlefont=cmb10 scaled\magstep2 

\font\headingfont=cmb10 at 12pt
%

\font\subheadfont=cmssi10 scaled\magstep1 
%








 %

\def\bra{\langle}
\def\ket{\rangle}
\def\xibar{\bar\xi}
\def\Wbar{\bar{W}}

\def\exp#1{ {\rm exp} \left\{ #1 \right\} }

\def\mean#1{{\left\langle #1 \right\rangle}}
\def\lambar{{\bar\lambda}}

\def\Wlarrow{\overleftarrow{W}\, }
\def\Wrarrow{\overrightarrow{W}\, }

\def\Gtilde{\tilde{G}}
\def\Itilde{\tilde{I}}
\def\Wtilde{\tilde{W}}
\def\boxtilde{\tilde{\block}}
\def\Pitilde{\tilde{\Pi}}

\def\blocktilde{\tilde{\block}}

\def\&{{\phantom a}}		



\phantom{}


\PrintVersionNumber   %


\sesquispace
\centerline{{\titlefont Scalar Field Theory on a Causal Set in Histories Form}\footnote{$^{^{\displaystyle\star}}$}%
{published in \journaldata{Journal of Physics: Conf. Ser.}{306}{012017}{2011}, 
\lbr \eprint{http://arxiv.org/abs/1107.0698} 
}}

\bigskip


\singlespace			        

\author{Rafael D. Sorkin}
\address
 {Perimeter Institute, 31 Caroline Street North, Waterloo ON, N2L 2Y5 Canada}
\furtheraddress
 {Department of Physics, Syracuse University, Syracuse, NY 13244-1130, U.S.A.}
\email{rsorkin@perimeterinstitute.ca}

\AbstractBegins                              
We recast into histories-based form a quantum field theory defined
earlier in operator language for a free scalar field on a background
causal set.  The resulting decoherence functional resembles that of the
continuum theory but the counterpart of the d'Alembertian operator
$\block$ is nonlocal and a generalized inverse of the discrete retarded
Green function.  We comment on the significance of this and we also
suggest how to include interactions.

\bigskip
%
\AbstractEnds                                

\bigskip



\sesquispace
\vskip -10pt

\section{1. Introduction}                                     
At this stage in the development of fundamental physics one does not
know whether anything like a quantum field will turn out to exist at the
Planck scale.
Once one abandons the continuum, a vector or a spinor is no longer a
particularly natural object, and in the context of causal set theory
further difficulties spring from the radical nonlocality inherent in a
discrete Lorentzian causal structure.  Such difficulties suggest that
the type of quantum field theory embodied in the so called standard
model will someday come to be seen as no more than a relatively low energy
description of underlying ``degrees of freedom'' of a completely
different nature.

Be that as it may, any attempt to push quantum field theory down to a
more fundamental, discrete level seems bound to be illuminating, and as
it happens, we now possess a theory that describes a free scalar field
propagating in a fixed, background causal set.\footnote{$^\star$}
{For some background on causal sets see 
 [1] [2] [3] [4] [5] [6]
 }
On the basis of this theory, one can hope
to understand better the influence of discreteness on such phenomena as
Hawking radiation or the propagation of wave packets over great
distances.
As formulated so far, however, the theory in question speaks the
language of operators and state-vectors.  It  describes the field by
attaching to each element $x$ of the causet $C$ an operator $\phi(x)$
acting in a Hilbert space built up from a ``ground state'' by the action
of creation operators $a^*$.  Although all this is very familiar, one
might wish also for a formulation based on ``histories'' (a history in
this case being a function $\phi:C\to\Reals$).\footnote{$^\dagger$}
{Sometimes people say ``classical history'' to emphasize that $\phi(x)$
 is a $c$-number and not an operator, but I will eschew that usage here
 in order to avoid any implication that $\phi$ must obey the classical
 equations of motion.}

Indeed, some of us regard histories-based formulations of quantum
theories as more basic and more satisfactory than operator formulations,
both for the purposes of quantum gravity and for the sake of
philosophical understanding.
How, for example, might one expect a dynamical theory of causal sets to
look?  The only way open to a ``law of motion'' for causets has so far
seemed to pass through something like a path integral (or more precisely
a decoherence functional), and one might say the same about quantum
gravity in general.  
Even with the causet limited to the background role of ``arena'', the
knowledge of how to write down a path integral for a scalar field on a
causet, would allow the concept of ``anhomomorphic coevent'' to be
brought to bear on the question of what sort of reality quantum field
theory is in fact describing [7].
In a more practical vein, one encounters the key question of how to
generalize from a free field to an interacting one, something which so
far has not been possible with the operator formulation.  As we will
see, a path integral formulation can suggest a very definite answer to
this question.

Of equal practical importance is the question of the classical limit.
Is it possible in this limit to express the equations of motion of a
scalar field in such a manner that solutions could be built up
iteratively given the values of the field in a sufficiently great
initial portion of the causet?  If it were, then computer simulations of
wave packet propagation through the causet would become possible.  On
the other hand, if the relevant field equation were to involve reference
to the distant future, such simulations would become much more
difficult, or even impossible.  In fact an equation of motion of exactly
the required, retarded sort has been proposed
[8], but in what sense 
can we claim that it represents the classical limit of our discrete
quantum field theory?

In this connection, the following difficulty appears to arise
[9].  Suppose that the dynamics of the scalar field is
to be defined through an action-integral (or sum in the case of the
causet), and that the integrand at point $x$ depends on the field in the
entire past of $x$.  When one varies $\phi(x)$ in such an expression,
the effect will be felt everywhere in the future of $x$.  The resulting
Euler-Lagrange equations will thus relate $\phi(x)$ not only to the
past but to the future.  How then could retarded equations of motion
result?  We will return to this issue in sections 4 and 5.

\section{2. Review of the continuum theory in operator form}  
Nothing could be more familiar than a free scalar field 
in Minkowski spacetime
together with
its vacuum-state and the concomitant Fock
representation of the field operators.  But the way in which these
things are ordinarily introduced, starting from the Klein-Gordon field
equations and the equal-time commutation relations, is not quite suitable
for transposition to the context of a discontinuous structure like the
causal set which not only does away with continuous time, but also
resists any notion of ``Cauchy data at a moment of time''.\footnote{$^\flat$}
{A causal set does admit the notion of {\it slice} as maximal antichain
 $\Sigma$, but such a subset unfortunately is not equipped to fill the
 role of an equal-time hypersurface, or more generally of a Cauchy
 surface in a globally hyperbolic spacetime, the most important
 difference being that the great majority of causal {\it links} will
 ``pass thru'' $\Sigma$ without meeting it at all;  
  that is, there will exist
 pairs of causet elements $x$, $y$ such that $x$ is to the past of
 $\Sigma$, $y$ to its future, and yet neither $\Sigma$ nor any other
 part of the causet contains an element causally intermediate between
 $x$ and $y$.  In this sense ``information can pass through $\Sigma$
 without registering on it'', the impossibility of which is precisely
 what characterizes a Cauchy surface in the continuum.}
Nor does the normally crucial distinction between positive and negative
frequency make much sense in a causal set, which insofar as it resembles
a spacetime at all, resembles much more a spacetime with curvature than
one without it.
For such reasons, it has proven more fruitful to rest the derivation of
the scalar quantum field theory as much as possible on the commutation
relations alone, or more precisely on the retarded Green function, in a
way that I will review here. [10]

By way of preparation, let us first review the theory of a ``gaussian''
noninteracting (massive or massless) real scalar field $\phi$ in flat
spacetime.\footnote{$^\star$}
{with a metric $\eta_{ab}$ of signature $(- + + +)$ and with
 $\block=\eta^{ab}\partial_a\partial_b$} 

\subsection {the usual story}
As this subject is commonly introduced, the input to the theory
comprises first of all the operator equations of motion in the form of
the Klein-Gordon equation, 
$(\block-m^2)\phi(x)$ = $0$.  
Supplementing these equations (plural since there is one for each $x$)
are the canonical commutation relations, expressed at a fixed moment of
time as
$[\phi(x),\partial\phi(y)/\partial{t}]=i\delta^{(3)}(x-y)$.  
One then expands $\phi$ in terms of (four-dimensional) plane waves and
identifies the (suitably normalized) coefficients of the
positive-frequency waves with annihilation operators $a(k)$.
Introducing, finally, the vacuum state $|0\ket$ via the conditions
$a(k)|0\ket=0$, one obtains a concrete family of field operators
$\phi(x)$ acting in a Hilbert space spanned by states of the form
$a^*(k_1)^{n_1} a^*(k_2)^{n_2} a^*(k_3)^{n_3} \cdots |0\ket$.  From an
algebraic standpoint, the equations of motion together with the
commutation relations define an abstract *-algebra generated by symbols
$\phi(x)$, while the choice of vacuum induces an irreducible
representation of this algebra in a particular sort of Hilbert space.
The full input to the theory is thus the field equations, the
commutation relations, and the choice of vacuum (although it is partly a
matter of opinion whether or not a change of representation or of vacuum
should be regarded as changing the theory per se.)

For our purposes, the most important thing to notice is the need for
consistency between the field equations and the commutation relations.
In the development we have just called to mind, this is ensured by the
fact that the latter are invariant under the time-evolution generated by
the former, but we can appreciate better what is going on if we express
the commutation relations in a manifestly covariant form, the so called
Peierls form.  (The Lorentz invariance of the theory then follows
trivially.)  To that end, let $G(x,y)=G^{ret}(x,y)$ be the retarded
Green function belonging to the wave-operator $\block-m^2$~:
$$
        (\block - \, m^2) G^{ret}(x,y) =  \delta^{(4)}(x-y) \ ,
$$
where $G$ is required to vanish unless $x\succ y$, meaning that $y$ is
within or on the past lightcone of $x$.  Further define
$$
           \Delta(x,y) = G(x,y) - G(y,x) \ ,   \eqno(1)
$$
which is just the difference between the retarded and advanced Green
functions, since (as in fact holds in any globally hyperbolic spacetime)
the advanced Green function coincides with the ``transpose'' of the
retarded one. The commutation relations then assume the remarkably
simple form,
$$
           [\phi(x), \phi(y)] = i \Delta(x,y) \ .
$$
If to this equation one applies the wave operator $(\block - \, m^2)$ (acting
on $x$), one trivially obtains zero on the left hand side.  The required
consistency then follows from the fact that the right hand side also
vanishes because $\Delta$ is the difference of two Green functions for 
$(\block - \, m^2)$. 
Notice here that the last three equations all make perfect sense, and
are valid, in any globally hyperbolic spacetime.  The resulting
definition of a quantum field theory accordingly goes through as well,
except that something is needed to replace the notion of positive
frequency if one wishes to define a vacuum state.

Now that we have a field theory in operator form, the next question is
how to render it into histories form.  For this there is a standard
procedure that I won't review here, because we will meet it again in
detail when we come to the case of the causal set.  Instead, let me just
quote the resulting decoherence functional, which by definition attaches
a complex amplitude to every {\it pair} of spacetime histories of the
scalar field.  (A pair of histories can be called a ``Schwinger
history'' in honor of the so called Schwinger-Keldysh version of the
path integral, which operates precisely with such pairs.)
To avert confusion between $\phi$ as a (``classical'') history and
$\phi$ as an operator-valued field, let me for now use the symbols $\xi$
and $\xibar$ to represent histories.   The decoherence functional 
$D(\xi,\xibar)$ is then given formally by the expression
$$
   \exp{i S(\xi)} \, \exp{i S(\xibar)}^*  \, \delta(\xi|\Sigma_T,\xibar|\Sigma_T)
   \eqno(2)
$$
where 
$S(\xi)=\half\int(-({\grad\xi})^2 -m^2\xi^2)$
plus an implicit further term that incorporates the effect of the
``initial state'', in this case the vacuum.   Here also, $\Sigma_T$ is
a spacelike future boundary at which the spacetime has been truncated
and the $\delta$-function involving it has the effect of forcing the
otherwise independent histories $\xi$ and $\xibar$ to share the same
restriction to this final boundary.
From (2), 
the decoherence functional extended to sets of histories and
the corresponding quantum measure can be obtained 
in the usual way [11].



For future reference let us  also define here the so called Wightman
or two-point function given by the expectation of $\phi(x)\phi(y)$
operators in the vacuum state:
$$
          W(x,y) = \bra \phi(x) \phi(y) \ket  \eqno(3)
$$
and let us note also that in the vacuum
$$
          W(x) := \bra \phi(x) \ket = 0 \ .  \eqno(4)
$$
Because the vacuum is a ``gaussian state'' the entire field theory is
effectively encapsulated in these two equations, a fact we will use
heavily in our construction of the decoherence functional in the causet
case.

\subsection {a different starting point}
The logic of the ``usual story'' that we have just rehearsed is
simplest if we present the commutation relations in their Peierls
form.  We might then summarize in part the steps we followed as: 
wave operator 
$\to$ $G^{ret}$ 
$\to$ $\Delta$ 
$\to$ $[\phi,\phi]$.  
But 
consistency 
between the wave-operator, 
$\block-m^2$~, 
and the
consequent commutation relations required, as we saw, that the transpose
of the retarded Green function also be a Green function for the same
wave operator.  Since no such relation is known in the causet case, it
turns out that one can get farther by changing the starting point
slightly from the equations of motion to the retarded Green function
$G^{ret}$.  From it we can derive the commutators in just the same way.
Indeed, one can pass freely from $G^{ret}$ to $\Delta$ and back by means
of equation (1) and the inverse relation
$G^{ret}(x,y)=\Delta(x,y)$, which holds when $x\succ y$.
(In equation form, $G^{ret}(x,y)=\theta(x,y)\Delta(x,y)$
where $\theta$ is the ``covariant Heaviside function''.)

But how to continue without the aid of any equations of motion?  The
surprising answer discovered implicitly in [10] is
that one can pass directly from the commutator function $\Delta$ to the
two-point function $W(x,y)$ by thinking of $\Delta(x,y)$ as a real skew
matrix $\Delta^{xy}$ and forming the positive part of the corresponding
Hermitian operator $i\Delta$.  Our new logic will therefore go as follows:
$G^{ret} \to \Delta \to W$.  
It would be difficult to imagine anything much simpler!

Notice in this connection that $W$ 
always is, 
in fact, 
a positive
(semidefinite) operator, as follows immediately from the fact that
$v^* W v\ge0$ for any complex vector $v$.  Indeed, positivity of the
Hilbert space inner product yields for any such $v_x$~, \ 
$\sum (v_x)^* W^{xy} v_y 
= \sum (v_x)^* \bra \phi^x \phi^y \ket v_y
= \bra X^\dagger X\ket 
= || X |0\ket ||^2 
\ge 0$~,
where $X=\sum v_x \phi^x$.
It's equally true that $\Wbar$, 
the complex conjugate matrix of $W$, 
is positive
semidefinite and that $i\Delta=W-\Wbar$, as follows from:
$(W-\Wbar)^{xy}=\bra \phi^x \phi^y\ket - \bar{\bra \phi^x \phi^y\ket}
=\bra \phi^x\phi^y\ket - {\bra \phi^y \phi^x\ket}
= \bra [\phi^x,\phi^y]\ket
= \bra i\Delta^{xy}\ket
= i\Delta^{xy}$~.  
Thus, $i\Delta$ 
is always the difference of two positive matrices $W$ and $\Wbar$.  
What our 
prescription 
for $W$
is adding to this fact is that these
operators have orthogonal support: 
$$
    W\Wbar=\Wbar W=0  \ .   \eqno(5)
$$
One might view this as a kind of ``ground-state condition'' imposed on
$W$ beyond what follows automatically from the fact that it is the
two-point function of a selfadjoint operator $\phi^x$.  
From
(5) follows immediately a simple matrix equation giving
$W$ in terms of $\Delta$  (and therefore in terms of $G$):
$$
          W = 1/2 (i\Delta + \sqrt{-\Delta^2})  \ .  \eqno(6)
$$
In practice one would often compute $\sqrt{-\Delta^2}$ by diagonalizing
$\Delta$.  Once this was done, one would obtain $W$ just by expanding
$i\Delta$ in terms of its eigenvectors and retaining the terms with
positive eigenvalues, and one could regard this as the practical meaning
of our prescription for obtaining $W$ from $\Delta$.

For future reference let me introduce the notation
$R=\half\sqrt{-\Delta^2}$, 
a manifestly real and positive matrix. 
Comparison with (6) then shows that 
$R$ is just the real part of $W$, 
whose decomposition into real and imaginary parts is
thus given by 
$$
   W = R + i\Delta/2  \ .                \eqno(7)
$$

It's worth noticing here that the definitions we have given yield a free
field theory with a distinguished ``ground state'' for any globally
hyperbolic spacetime or region of spacetime 
(modulo certain technical questions about convergence that arise when
the region's volume is infinite.)
Notice also that nowhere do these definitions
refer to any notion of positive frequency.  In Minkowski spacetime, the
new prescription can easily be seen to reproduce the usual vacuum (it
more or less had to by Lorentz invariance).  In this sense it
generalizes the Minkowski vacuum to the case of arbitrary curvature.

Perhaps also it's also worth mentioning a possible further generalization
that would tie $W$ less closely to $G$.  In order to reproduce correctly
the commutation relations, it is necessary only that the imaginary part
of $W$ coincide with $\Delta/2$.  
It is also necessary for mathematical consistency that $W$ be positive.
Perhaps other conditions are needed as well, 
but it is clearly not necessary 
that $W$ simply be the positive part of $i\Delta$.  
One could thus consider relaxing condition
(5) as well as the assumption that $\bra\phi\ket=0$.  
As will be seen, neither generalization would require major changes to our
discussion below of the decoherence functional on a causal set.

\section{3. Scalar field theory on a causet in operator form} 
Let us now leave the continuum behind and turn to the case of a causal set.
Henceforth our discussion will be in the context of a fixed, finite
causal set $C$, and it will assume further that a fixed ``retarded Green
function'' has been adopted, which I'll write as either $G(x,y)$ 
or $G^{xy}$~, 
where $x$ and $y$ vary over the elements of $C$.
In some of the most important cases a convincing candidate for $G$ is
known and has been tested in practice to a greater or lesser extent.
These cases include zero mass in a causet intended to approximate an arbitrary
2-dimensional spacetime, as well as arbitrary mass in a causet intended to
approximate a flat spacetime of dimension either 2 or 4. 
(See [12].)
In addition, the ``retarded d'Alembertians'' of
[8]
and [13] could be
inverted to produce candidate Green functions in more general causets,
but this type of prescription has not yet been put to a test.

Starting from $G$ and following the same steps as delineated above
yields then a scalar field theory for the given causet, complete with
operators $\phi(x)$, vacuum-state $|0\ket$, and Wightman function
$W(x,y)=\bra \phi(x) \phi(y)\ket$, where all expectations are taken with
respect to the state $|0\ket$.

As we have already noted, the fact that this theory-cum-state is in some
sense gaussian means that all its consequences can be recovered starting
from $W$ alone.  It might be interesting to tease out better in exactly
what sense the word ``gaussian'' can be applied to our causet theory.
Having no Cauchy surfaces, it also has no straightforward
Schr{\"o}dinger representation, and therefore no vacuum wave function
whose exponential form could be held up as the meaning of the word.  On
the other hand, the fact that the $\phi^x$ are linear combinations of
raising and lowering operators implies that Wick's theorem will work in
the expected manner, and this is all that we will need.  Indeed, we will
need only the following consequence of Wick's theorem, where $\Phi$
stands for any linear combination of the field operators $\phi(x)$.
$$
   \mean{\exp{\Phi}} = \exp{ {\mean{\Phi\Phi}\over 2} } 
   \eqno(8)
$$
In view of equation (20) of the appendix, the following calculation
demonstrates this identity.
$$
  \bra \exp{\Phi} \ket
 = \sum_n { \mean{{\Phi}^n} \over n! }
 = \sum_n { \mean{{\Phi}^{2n}} \over (2n)!}
 = \sum_n  {(2n-1)!! \mean{\Phi\Phi}^n \over (2n)!}  
 = \sum_n  {1 \over n!}  {\mean{\Phi\Phi}^n \over 2^n}
$$

This seems a good place to mention a way of thinking about $W$ that is
more geometrical in nature than the characterizations we have
encountered so far and that involves only real, as opposed to complex,
vector spaces.
Namely $W$ can be represented geometrically as a family of mutually
orthogonal, ``axially weighted 2-planes'' in the real vector space
$\Reals^N$, where $N$ is the cardinality of the underlying causet, 
these being a family of pairs $(\varpi,\lambda)$ where $\varpi$ is an
oriented 2-plane in $\Reals^N$ and $\lambda$ is a strictly positive real
number corresponding to one of the eigenvalues of $i\Delta$.
This one can see by splitting the eigenvectors of $i\Delta$ into their
real and imaginary parts, or equivalently by decomposing $\Delta$ as a sum
of terms $a\wedge b$ with $a,b\in\Reals^N$ being orthogonal vectors.
The latter vectors then provide the basis of a singular-value
decomposition of the real matrix $\Delta$.  Thanks to these
relationships, one can construct $W$ directly from $\Delta$ by forming
the singular value decomposition of the latter, thereby economizing ---
in computer simulations --- on memory and CPU-time, since only real
numbers need be involved.

We are almost ready to turn our efforts to producing a histories version
of our theory, but first, perhaps a simple example of the
scheme $G\to\Delta\to W$ would be in order.  Consider then a causet of
only two elements, $e_0\prec e_1$, making up a 2-chain.  (Given the
interpretation of $W$ as a collection of weighted 2-planes, we could in
this very simple case pass immediately to a unique $W$ without bothering
with $G$ and $\Delta$, but that would not illustrate the general
situation.)  That $G$ is retarded means precisely that $G^{01}=0$, but
since only off-diagonal terms survive the antisymmetrization in
(1), we might as well suppose that the diagonal of $G$ vanishes
as well.  Up to sign and normalization, $G$ is then unique:
$$
           G = \pmatrix{0 & 0 \cr 1 & 0}  \ ,
$$
from which there result immediately
$\Delta = \pmatrix{0 & -1 \cr 1 & 0}$
and
$$
   i\Delta = \pmatrix{0 & -i \cr i & 0}=\sigma_2
$$
(the Pauli matrix).  Substituting this into (6) then furnishes
$W$ as
$$
   W = \half(1+\sigma_2) = \half \pmatrix{1 & -i \cr i & 1}  \ ,
$$
i.e. $2R$ in (7) is simply the identity matrix.
We could also have arrived at the same result by diagonalizing
$i\Delta$ explicitly as $\sigma_2=u u^\dagger
- \bar{u} \bar{u}^\dagger$, where $u=1/\sqrt{2}\pmatrix{1 \cr i}$, and
then discarding the negative term.
Notice finally that $W$ correctly satisfies the ``ground-state
condition'' (5).  Given that $W$ is hermitian,
(5) is just the assertion that $W\Wtilde=0$, tilde
denoting the matrix transpose.  Alternatively expressed,
this says that
 the dot-product
(the bilinear one, not the hermitian one) of any row (resp. column) with
itself or with any other row must vanish, and this is true by
inspection.

\section{4. Scalar field theory on a causet in histories form}
Henceforth it will be convenient to redefine $G=G^{jk}$ to be the
retarded Green function {\it with its diagonal set to zero}.  
Since $\Delta$ is $G$ minus its transpose, it is unaffected by this
modification, and since $W$ was defined through $\Delta$, it is not
affected either.
Equivalently, we could take $W$ as our starting point and then
simply define $G$ to be twice the retarded half of the imaginary part of
$W$.  Either way its clear that for purposes of this paper, there's no
harm in dropping diag$(G)$.  But in a larger context this change would
need to be borne in mind.  For example, if we had obtained $G$ by
inverting some retarded D'alembertian operator for the causet, then we
could not recover the latter from $G$ without restoring its diagonal.

By a histories-based formulation of a field theory, one means 
(for a real scalar field)
a formulation that works directly with field configurations
$\phi:C\to\Reals$ (``histories''), and avoids reference to field
operators or state-vectors, except possibly as auxiliary technical
devices [14].  In practical terms, this means a path-integral
formulation, although one might hope that more general frameworks will
one day be available.\footnote{$^\dagger$}
{perhaps obtained by
 freeing the concept of preclusion from its dependence on the
 notion of quantum measure and measure-zero, or perhaps even 
 by freeing
 anhomomorphic inference from its dependence on preclusion,
 cf. [7]} 
In order that a framework based on path-integrals be self-contained, moreover, it
seems necessary to assign amplitudes to Schwinger histories rather
than to individual histories, as the more familiar path-integral does.  
That is to say that one needs to express the dynamics in terms of a
{\it{decoherence functional}} $D$, from which the corresponding quantal
measure 
[15]
can then be computed.  Since the definitions of these
things have been presented many times in the literature, I'll not
repeat them here.  Rather, I'll begin by writing down the
expression (an expectation of a product of projectors)
that one needs to evaluate in order to recover the
decoherence functional from a theory expressed in the operator
language.  

In line with the definitions laid down above,
let  $\xi$ and $\xibar$ be two (completely independent) histories, 
each being specified by a list of real numbers $\xi^x$, 
one for each element $x$ of the causet.
The complex number $D(\xi,\xibar)$ is then given by the equation
$$
   D(\xi,\xibar) =
   \mean{
    \delta(\phi^1-\xibar^1) \delta(\phi^2-\xibar^2) \cdots \delta(\phi^N-\xibar^N)
    \delta(\phi^N-\xi^N)  \cdots  \delta(\phi^2-\xi^2)  \delta(\phi^1-\xi^1) 
    }
\eqno(9)
$$
Here $\phi^j=\phi(x_j)$, $\xi^j=\xi(x_j)$, etc, where $j=1\cdots N$ is
any {\it natural labeling} of $C$.  In other words the elements
$x\in{C}$ of the causet must be labeled so that no element with a
smaller label ever temporally follows an element with a higher label:
$x_j\prec x_k\implies j<k$.  Such a labeling always exists, and any two
natural labelings are guaranteed to produce the same result for
$D(\xi,\xibar)$ because $\Delta(x,y)$,
and hence $[\phi(x),\phi(y)]$,
automatically vanishes when $x$
and $y$ are causally unrelated (``spacelike'').  Notice that $D$ is
normalized such that $\int d^N\xi d^N\xibar D(\xi,\xibar)=1$. 

Re-expressing the $\delta$-functions as integrals puts
(9) into the form
$$
   \int\limits_{-\infty}^{\infty} 
   {d\lambda_1\over 2\pi} {d\lambda_2\over 2\pi}
   \cdots
   {d\lambar_2\over 2\pi} {d\lambar_1\over 2\pi} 
   e^{-i\lambar_1\xibar^1 -i\lambar_2\xibar^2\cdots -i\lambda_2\xi^2 -i\lambda_1\xi^1}
   \mean{e^{i\lambar_1\phi^1} e^{i\lambar_2\phi^2}\cdots e^{i\lambda_2\phi^2} e^{i\lambda_1\phi^1}}
  \ ,
 \eqno(10)
$$
which represents $D(\xi,\xibar)$ as the Fourier transform of what one
might call its ``non-commuta\-tive characteristic function'' in analogy
with the concept of characteristic function that figures in ordinary
probability theory.  
(Here again the real numbers $\lambda_j$ and $\lambar_j$ are independent
parameters of the Fourier transform.  The bar is not being used to
denote complex conjugation.)

Before plunging into the evaluation of this expression for the
decoherence functional, let us evaluate the simpler expression which
results from integrating out the $\xibar$ and all but one of the
variables $\xi=\xi(x)$ in $D(\xi,\xibar)$, which mathematically is
analogous to a marginal probability density for the remaining variable
$\phi(x)$.  After doing the integrals, we are left simply with
$$
   \int {d\lambda\over 2\pi} \exp{-i\lambda\xi} \mean{\exp{i\lambda\phi(x)}}
  \ ,
$$
which in light of the identity (8) turns into
$$
    \int {d\lambda\over 2\pi} \exp{-w^2\lambda^2/2-i\lambda\xi}
     = {1\over\sqrt{2\pi w}} e^{-\xi^2\over2w}  
\ ,
$$
where I have written $\xi$ for $\xi(x)$ and $w$ for $\mean{\phi(x)\phi(x)}$.
Unsurprisingly for a free field, 
we just obtain
a gaussian.

The next simplest case, which already illustrates most of the
complications, results from integrating out all but two of the
field-values, and yields an integral of the form
$$
   \int {d\alfa\over 2\pi} {d\beta\over 2\pi}
   \exp{-i\alfa\xi-i\beta\eta}
   \mean{\exp{i\alfa\phi(x)}\exp{i\beta\phi(y)}}   
  \ .
$$
We can again avail ourselves of (8),
but only after bringing  to bear the well known identity
$$
    e^A e^B = e^{A+B} e^{\half[A,B]}  \ ,  \eqno(11)
$$
which holds whenever the commutator $[A,B]$ is a $c$-number.
In virtue of 
(11)
we find 
$$
  \exp{i\alfa\phi(x)} \exp{i\beta\phi(y)} =
  \exp{i\alfa\phi(x)+i\beta\phi(y) - \half\alfa\beta[\phi(x),\phi(y)]}
$$
which can also be written as 
$\exp{i\alfa\phi(x)+i\beta\phi(y) - \half\alfa\beta(W^{xy}-W^{yx})}$
because 
$$
  [\phi(x),\phi(y)] = \phi(x)\phi(y)-\phi(y)\phi(x)
  = \mean{\phi(x)\phi(y)-\phi(y)\phi(x)}
  = W^{xy}-W^{yx}
  \ ,
$$
the original commutator being a $c$-number.
Hence,
$$
\eqalign{
  & \mean{\exp{i\alfa\phi(x)}\exp{i\beta\phi(y)}} \cr
  &=
  \exp{-\half  \left(
  \mean{(\alfa \phi(x)+\beta \phi(y))^2}
   +\alfa\beta(W^{xy}-W^{yx})  \right)} \cr
  &=
  \exp{-\half  \left(\alfa^2 W^{xx} + \beta^2 W^{yy} +\alfa\beta W^{xy}+\alfa\beta W^{yx}
     +\alfa\beta W^{xy}-\alfa\beta W^{yx} \right)}  \cr
  &=
  \exp{-\half  \left(\alfa^2 W^{xx} + \beta^2 W^{yy} +2\alfa\beta W^{xy} \right)} \cr
  &=
  \exp{-\half\left(
        \pmatrix{\alfa & \beta} 
        \pmatrix{W^{xx} & W^{xy} \cr 
                 W^{xy} & W^{yy}}
        \pmatrix{\alfa \cr \beta} 
            \right)}   }
$$
The important thing to notice here is that the order of the indices
in  $W^{xy}$ copies the order of $x$ and $y$ in the original factors.

Following this pattern, it is straightforward to evaluate the quantity 
within angle brackets in (10).  
The result is
$$
  \exp{-\half\left(
   \Wrarrow^{jk}\lambar_j\lambar_k +
   \Wlarrow^{jk}\lambda_j\lambda_k +
    2W^{jk}\lambar_j\lambda_k
            \right)}   
\eqno(12)
$$ 
where 
$\Wrarrow^{jk}$ denotes 
$W^{jk}$ if $j\le k$ and 
$W^{kj}$ if $k\le{}j$, 
for example
$\Wrarrow^{23}=\Wrarrow^{32}=W^{23}$; 
and
$\Wlarrow$ follows the reversed convention.  
(That the arrows in the first and second terms point in opposite
 directions merely reflects the opposite ordering of the causet elements
 in
 the two halves of the Schwinger history.)
Notice in connection with (12) that the matrices $\Wlarrow$ and
$\Wrarrow$ are complex conjugates of each other, and recall also that
$W$ itself is positive semidefinite.

Let us observe in passing that integrating out the $\xibar$ variables
from $D(\xi,\xibar)$ would have the effect of setting $\lambar=0$ in
(12), leaving in the exponent only 
$-\half\Wlarrow^{jk}\lambda_j\lambda_k$, which involves only the
time-ordered Wightman function, 
or equivalently the ``Feynman Green function''. 


Now since $D(\xi,\xibar)$ is nothing more than the Fourier transform of
(12), its evaluation would be relatively carefree, were it not for
the fact that the quadratic form 
in
(12) is not invertible.  This will prevent $D(\xi,\xibar)$ from
assuming the pure gaussian form that it would otherwise have had, and
that one might naively have expected from a free field theory.  Instead,
it will be the product of a gaussian with a $\delta$-function that will
enforce certain ``constraints'' on its arguments.  

The further manipulations to be done will become more transparent if at
this point we switch to a matrix notation, and I'll 
continue to
use a tilde to
denote transpose.  Then for example,  we can re-express (12) as
$\exp{-\half Q}$ where
$$
      Q = \lambda \, \Wlarrow\, \lambda + \lambar \, \Wrarrow\,\lambar + 2\lambar\, W\,\lambda \ .
$$
Let us now express $Q$ directly in terms of $G$ and $R$ using that,
according to its definition,
$W=R+i\Delta/2=R+(i/2)(G-\Gtilde)$, and noting also
that
${\overleftarrow{R}\, }={\overrightarrow{R}\, }=R$
and
${\overleftarrow{\Delta}\, }=-{\overrightarrow{\Delta}\, }=G+\Gtilde$,
whence
$$
  \Wlarrow = \overleftarrow{R+i\Delta/2}=R+i\overleftarrow\Delta/2=R+(i/2)(G+\Gtilde)
$$
whence  similarly $\Wrarrow=R-(i/2)(G+\Gtilde)$.  
Expanding $\Wlarrow$ this way yields
$$
       \lambda \, \Wlarrow \, \lambda = 
       \lambda (R+(i/2)(G+\Gtilde)) \lambda = 
       \lambda  R \lambda + i \lambda G \lambda
$$
with similar expansions for  
$\lambar\Wrarrow\lambar$ and   $2\lambar W\lambda$.
Adding these together then reveals that $Q$ is expressed most simply
in terms of sum and difference variables for $\lambda$ and $\lambar$.
$$
   Q = (\lambda+\lambar)R(\lambda+\lambar) + i (\lambda+\lambar)G(\lambda-\lambar) \ .
$$
Its fourier transform can then be expressed in terms of
the  corresponding sum and difference variables for $\xi$ and $\xibar$.
To that end, let us define 
$$
  K = \lambda+\lambar \ ,\  k = \half (\lambda-\lambar) \ ,\ \ 
  \phi = \half(\xi+\xibar) \ ,\  \varphi = \xi-\xibar
$$
Here $\phi$, the mean of the two halves of the Schwinger history, is
what is sometimes called ``the classical field'' because 
$\xi=\xibar$ in the classical limit, while $\varphi$
represents in some sense the deviation from classicality.\footnote{$^\flat$}
{Please don't confuse $\phi$ as defined here with the same symbol used
 earlier for the field {\it operator}.
}

With these definitions, 
$\lambda\xi+\lambar\xibar=K \phi + k\varphi$, 
and the fourier transform yielding $D(\xi,\xibar)$ turns into
$$
   \int d^NK\, d^Nk\; \exp{-iK\phi-ik\varphi - \half KRK - i KGk} \eqno(13)
$$
where I've left out an overall normalization which needn't be carried
along because it can be recovered at the end from the condition
$\int d^N\xi \, d^N\xibar \, D(\xi,\xibar)= D(\Omega,\Omega)=1$.

Evidently the conjugate variable $k$ occurs only linearly in the
exponent of (13), in the combination
$k\varphi+KGk=k(\Gtilde K+\varphi)$;
the $dk$ integral therefore produces the
$\delta$-function I mentioned earlier, which constrains $\varphi$ to be in
the image of $\Gtilde$, the ``advanced Green function''.  Equivalently,
it constrains $\varphi$ to be orthogonal to 
$\ker(G)=\SetOf{K}{GK=0}$~.  
We now have
$$
  D(\xi,\xibar) = (const) \int d^NK \; \delta(\Gtilde K + \varphi) \;
                \exp{ -\half KRK -iK\phi }  \ .  \eqno(14)
$$

When $\varphi$ is in the image of $\Gtilde$, as required for $D$
to be nonzero, (14) has the form of a gaussian integral over an
affine subspace of $\Reals^N$, namely the space 
of solutions 
to the
equation $\Gtilde K+\varphi=0$.  To convert (14) into an
integral over a full vector space, let $I:V\to\Reals^N$ be the inclusion
map of 
$V=\ker\Gtilde$ 
into $\Reals^N$ and let $K_0$ be any solution of the equation 
$\Gtilde K_0=-\varphi$~,
so that the space over which we have to integrate is $K_0+V$.
(Concretely we can represent $I$ by a matrix $I_j^{\&\alfa}$, where the
vectors $I_j^{\&\alfa}$ for fixed $\alfa$ furnish a basis for $V$.)
The substitution of $K_0+K$ for $K$ then yields
our integral in the form
$$
\eqalign{
    & \int_{K\in K_0+V} \, d^{N'}K  \; \exp{-\half KRK - iK\phi} \cr
    & =  \int_{K\in V} \, d^{N'}K  \; \exp{-\half (K+K_0)R(K+K_0) - i(K+K_0)\phi} \cr
    & =  \exp{-\half K_0RK_0-iK_0 \phi}\int_{K\in V} \, d^{N'}K  \; \exp{-\half KRK - K(RK_0+i\phi)}
}
$$
where $N'=\dim\ker\Gtilde$ is the dimensionality of $V$.   
Then
with the
aid of the matrix $I$ 
we obtain 
an integral over $\Reals^{N'}$ by
putting $Iv$ in place of $K$~:
$$
   \exp{-\half K_0RK_0-iK_0 \phi} 
   \int\,d^{N'}v  \; 
   \exp{-\half v\Itilde RIv - (K_0R+i\phi)Iv}
   \ . \eqno(15)
$$
The result is now an ordinary gaussian integral over $\Reals^{N'}$ of the form
$\int\exp{-\half vAv + Bv}$ with $A=\Itilde R I$ and $B=(K_0R+i\phi)I$.  
Under the assumption that $A$ is invertible,
this integral evaluates to (a constant multiple of) 
$\exp{-\half B (1/A)B}$~.  
When expanded out, the exponent here consists of three terms:
$$
   \phi \, I(\Itilde R I)^{-1} \Itilde \, \phi \ ,  \eqno(16-1)
$$
$$
       -i K_0 \, (1-RI(\Itilde R I)^{-1} \Itilde) \, \phi \ , \eqno(16-2)
$$
$$
          -\half K_0 \, (R - R I (\Itilde R I)^{-1} \Itilde R) \, K_0 \ . \eqno(16-3)
$$
We now have the decoherence functional in hand, except that (16)
is expressed in terms of the ambiguous constant $K_0$ rather than
$\varphi$.  For consistency, this apparent dependence on $K_0$ must be
illusory, i.e. (16) must be independent of $K_0$ for fixed
$\varphi$.

In considering (16),
let us observe first of all that
one would expect $I$ to be ``small'' since one would expect
$\dim\ker\Gtilde$ to be small compared to $N$, the number of elements in
our causet.  It can't vanish entirely, given that $G$ is strictly
retarded --- now that we've removed its diagonal --- but an analogy with the
continuum might suggest that it would receive contributions only from
the ``boundary region'' of the causet.  To the extent that $I$ actually
can be neglected, (14) could have been evaluated without doing any
integrations at all, as the $\delta$-function therein sets $K$ to
$-\Gtilde^{-1}\varphi$, yielding
$D=\exp{-\half\varphi(G^{-1}R\Gtilde^{-1})\varphi + i \varphi G^{-1}\phi}$.
Now a natural symbol to represent $G^{-1}$ would be $\block$~, since in
the continuum $G$ is in some sense the inverse of $\block$.  In the
present setting we can also appreciate that $G^{-1}$ (if it actually
existed) would be retarded too because $G$ itself is retarded.  One
might then refer to $\block=G^{-1}$ as a ``retarded d'Alembertian'' and
its transpose $\boxtilde$ as an ``advanced d'Alembertian''.  With this
nomenclature we would have $D=\exp{-\half\varphi(\block R\boxtilde)\varphi
+ i \varphi \block \phi}$~.  Moreover, in the continuum, $\block R= \Re (\block
W)=0$, therefore one might also expect $-\half\varphi(\block
R\boxtilde)\varphi$ to be small in the causal set.
In that case the whole of the decoherence functional would simplify to
$\exp{i \varphi \block \phi}$~.
Remarkably, this is exactly the decoherence functional of the continuum
theory, as an integration by parts reveals, given that $\varphi=0$ on
the future boundary.  More precisely, the continuum decoherence
functional is this expression together with 
the $\delta$-function in (2)
that
mandates the just-mentioned equality between $\xi$ and $\xibar$ on the
future boundary, cf. [11].  Such a
$\delta$-function is present implicitly in (14) and explicitly in
(9), but I don't know how to extract it as such from a
well-defined continuum limit of the discrete theory.  We'll return to
the constraints on $\varphi$ shortly.

It is useful at this point to distinguish conceptually between the space
$\Reals^N$ of field-configurations and the space $\Reals^N$ of
configurations of the dual variables $K$, $\lambda$, etc.  Since we have
already employed the symbol $\Omega$ for the former, we might as well
denote the latter by $\Omega^*$~, using a star to denote dual space.
I'll also write $V$ for $\Reals^{N'}$
considered as $\ker\Gtilde$ considered as a separate vector space.  Then
$I:V\to\Omega^*\,$, \  
$\Itilde:\Omega\to V^*\,$, \ 
$G:\Omega^*\to\Omega\,$, \ 
$\Gtilde:\Omega^*\to\Omega\,$, \ 
etc.

Let us return now to the full decoherence functional as given by
(16).  Its second member can be written as $-iK_0(1-\Pi)\phi$ if
we define $\Pi$ to be the operator
$\Pi=R I(\Itilde R I)^{-1}\Itilde$~.  Here, of course, we continue to
assume that $(\Itilde R I)^{-1}$ exists.  From its definition it is
trivial to check that $\Pi$, though not hermitian, is a projection in
the sense that $\Pi^2=\Pi$; and also that
$$
         \Itilde \; \Pi = \Itilde \ , \quad \Pitilde \; I = I  \ .  \eqno(17)
$$
Now the term by which $K_0=-\Gtilde^{-1}\varphi$ is ambiguous is by
definition $Iv$ for arbitrary $v\in V$.  Hence the ambiguity in
$-K_0(1-\Pi)\phi$ is precisely
$(Iv)(1-\Pi)\phi=v\Itilde(1-\Pi)\phi=v(\Itilde-\Itilde\Pi)\phi$, which
vanishes by (17).  Consequently, it is justified to write 
(16-2) simply as 
$i(\Gtilde^{-1}\varphi)(1-\Pi)\phi=i\varphi G^{-1} (1-\Pi)\phi$~. 
In view of these considerations, it is natural to define
$$
           \block = G^{-1} (1-\Pi)     \ , \eqno(18)
$$
meaning that $\block$ solves the equation $G\block=1-\Pi$.
We have just seen that $\block\phi$ is unambiguous when contracted with
any $\varphi\in\im(\Gtilde)$.  Thus $\block$ itself is uniquely defined if
we construe it as a map from $\Omega$ to $(\im\Gtilde)^*$, the dual
vectorspace to $(\im\Gtilde)$~: in symbols,
$\block:\Omega\to(\im\Gtilde)^*\part\Omega^*$~.
Regarded however as mapping $\Omega$ to all of $\Omega^*$, $\block$ is
ambiguous by the addition of any linear operator with image in 
$(\im\Gtilde)^0=\ker G$~.  
A nice idea at first sight would be to take
advantage of this ambiguity to render $\block$ fully retarded, but
unfortunately that is not possible.
Reasoning similar to the above also proves that
$G\block G=G$, 
that
$\im\Pitilde=\im I = \ker \Gtilde$,
that
$\ker\Pi=\im G$,
and therefore that
$\im(1-\Pi)=\im G$.
Also that
$\Pi R = R\Pitilde = RI(\Itilde RI)^{-1}\Itilde R$~.


So far, we have been occupied primarily with the analysis of (16-2), 
but we also need to deal briefly with (16-1) and (16-3). 
The former is complete as it stands, but the latter requires to be
expressed in terms of $\varphi$, just as we did for (16-2).
To that end, rewrite (16-3) as
$ -\half K_0(1-\Pi)R K_0 
= -\half K_0(1-\Pi)^2 R K_0 
= -\half K_0(1-\Pi) R (1-\Pitilde) K_0
$ 
and observe that the ambiguity of adding a vector $Iv$ to $K_0$ drops
out just as before because $(1-\Pitilde)I=0$.  Continuing on, and replacing $K_0$ by
$-\Gtilde^{-1}\varphi$, we obtain with the aid of equation (18),
$ -\half \varphi G^{-1} (1-\Pi) R (1-\Pitilde) \Gtilde^{-1} \varphi
= -\half \varphi \block (1-\Pi) R \blocktilde \varphi$,
which can also be written as
$-\half (\blocktilde \varphi) R (\blocktilde \varphi)$.  
Collecting our
three terms then yields in total
$$
     D(\xi,\xibar) = (constant) \;  \exp{
                   i\varphi\block\phi
                   -\half (\Itilde\phi)(\Itilde RI)^{-1}(\Itilde\phi)
                   -\half (\blocktilde \varphi) R (\blocktilde \varphi) }
   \eqno(19)
$$ 
In interpreting this formula, one should bear in mind that it includes
an implicit delta-function\footnote{$^\star$}
{Since $\im\Gtilde=\orthog\ker{}G$, we could make the delta-function
explicit by selecting a basis of (co)vectors $v$ for the kernel of $G$
and forming the product over these $v$ of $\delta(v\cdot\varphi)$}
that restricts $\varphi$ to lie in the image of $\Gtilde$.  
That is, $D(\xi,\xibar)$ vanishes unless $\xi$ and $\xibar$ differ by an
element of $\im\Gtilde$. 
If we remember also that with our definitions, the diagonal elements of
$G$ all vanish,
then we see immediately
that every function in $\im\Gtilde$ vanishes on
the maximal elements of our causal set $C$,
which implies in particular
that (just as in (2)) $\xi$ and $\xibar$ must agree on these
elements. 
In general though, there will arise additional linear
relations among the $\varphi^j$ that lack any obvious continuum analog.

In view of the rather intricate derivation we've just been through
it seems worth recording here the much simpler expression that results
from integrating out the $\xibar$ variables in $D(\xi,\xibar)$ or simply
from returning to (9) and deleting the factors involving
$\xibar$.  One sees immediately from (12) that --- provided that
$\Wlarrow$ is invertible --- the result will be
$$
       D(\xi,\Omega) = \int d^N\xibar \, D(\xi,\xibar) = 
     (const) \; \exp{{i \over 2} \xi \block_F \xi } 
$$ 
where I've written $\block_F$ for $i\Wlarrow^{ -1}$, a notation that seems
natural because $-i\Wlarrow$ corresponds to the Feynman Green function of
the continuum theory.

\section{5. Consequences and questions}                       
Perhaps the most important conclusion from the above work is that the
equations of motion for the ``classical'' field $\phi$ are essentially a
retarded non-local version of the continuum equations $\block\phi=0$.
Equations of precisely this sort were proposed in 
[8], 
but until now
there has seemed to be no sound reason for preferring retarded equations
over, for example, the time-reverse or some combination of the two.
Granted that retardation greatly facilitates computer solution, but that
would be more an opportunistic reason than one of principle.
Retardation also agrees better with the notion that physics in the
classical regime, even if it were to be non-local, would still be
``causal'',
but
here we have reached a similar conclusion in a far more convincing
manner!


But why should one refer to $\block\phi=0$ as the classical equations of
motion?  Expressed through the decoherence functional, classicality
amounts to the vanishing of $\varphi$, i.e. to the equality of $\xi$
with $\xibar$ (in which case both coincide with $\phi$).  But if we vary
$\varphi$ in (19) and then set $\varphi=0$, the result is precisely
$\block\phi=0$.  Further support for this point of view comes from the
fact, already alluded to, that the second and third terms in
(19) seem likely to be relatively negligible, having no further
significance than either initial-conditions or small corrections.  In
the case of the second term specifically (the term in $\phi$), its
character as a kind of initial condition or ``state'' can be perceived
fairly clearly, because it depends on $\phi$ only through $\Itilde\phi$,
which is a combination of contractions of $\phi$ with vectors in
$\ker\Gtilde$.  But since $\Gtilde$ is ``future-looking'' it will
annihilate any function with support on the minimal level of $C$.
Plausibly then, its entire kernel is comprised of functions supported
near this bottom level, whence $\Itilde\phi$ would depend only on such
initial values of $\phi$. 

Of course if we vary all the arguments in (19), we obtain more
equations, and taken together they tell us more than just
$\block\phi=0$.  In fact they constrain $\phi$ itself to be zero (along
with $\varphi$).  This however only reflects the fact that our action
(or rather decoherence functional) implicitly includes a specification
of ``initial conditions'', something like an ``initial state
wave-function'' in the continuum theory.  In fact exactly the same thing
happens in that case: a unique ``classical'' solution is picked out.  In
the present context it's unremarkable that the corresponding solution is
$\phi=0$, because our construction of the decoherence functional assumed
early on that 
(4) was true.


We might also ask how an apparently time-symmetric scheme has given rise
to a very time-asymmetric set of classical equations.  To the extent
that we can regard $\block$ as retarded, the combination
$i\varphi\block\phi$ is entirely retarded in the ``classical component''
$\phi$ and entirely advanced in the 
``non-classical component'' 
$\varphi$.  
Just how
this has come about seems mysterious, but logically the asymmetry had to
originate in the order chosen for the factors in (9).
Perhaps a careful tracing through of the steps leading from there to
(19) would shed more light on the question.

Of course the operator $\block$ is not fully retarded but only
approximately so (if our plausibility arguments are valid).  Should this
worry us?  I'm not sure, but one reason for equanimity at this point
springs from the observation that a dynamics taking place in a frozen
causal set can only be a partial reflection of full quantum gravity.
Really the causet itself needs to be included in the decoherence
functional.  Not having done this we have restricted its future
development in an unphysical manner, and thereby we have imposed a sort
of future boundary condition.  It would be no surprise if such a
condition were to be reflected to some extent in the dynamics of the
variables (namely the scalar field) that we have left free.
(Whether we can be equally happy with an anti-causal equation for
$\varphi$ is another question however.)

A question might be raised here about our assumption that the matrix
$\Itilde RI$ is invertible.  How well does this hold up in practice?
Quite well if the few numerical experiments that I've done are
indicative, but so far they have been limited in scope.  An interesting
exception, however, springs from the existence of so called non-Hegelian
elements or pairs within the causal set $C$, viz. pairs of incomparable
elements $x$ and $y$ which share the same relation to every third
element $z\in{C}$.  Such elements cannot be distinguished from each
other by the causal structure, and in consequence the ``difference
function'' defined by $f(x)=+1$, $f(y)=-1$, and $f(z)=0$ otherwise, will
be in the kernel of $G$, $\Delta$, etc.  In particular it will be in the
kernel of $\Itilde RI$, rendering the latter non-invertible.  But the
problem this might seem to raise solves itself.  Indeed $x$ and $y$ are
in some sense not two distinct elements at all, and the theory seems to
know this, as one can see by returning to equation (13) or
(14) and noticing that $f$ is a degenerate direction for the
quadratic form involved.  Consequently the resulting decoherence
functional will vanish unless $\xi(x)=\xi(y)$ and $\xibar(x)=\xibar(y)$.
The theory thus lives in effect on the quotient causet $C'$ obtained by
identifying $x$ with $y$, and one might just as well eliminate $y$ (say)
at the outset and continue in this manner until no further non-Hegelian
elements remain.

Many more questions could be raised at this point, but let me just
conclude with two of a more general nature, the first being this.  All
our work herein has presupposed a fixed, background causal set $C$ with
a finite number of elements.  But in reality, one expects the causet to
be growing and infinite, or at least potentially infinite toward the
future.  We cannot actually handle dynamical growth without setting up a
full quantum gravity theory, but we can certainly imagine $C$ being
infinite toward the future.  In that case our theory herein would
correspond to only some initial portion of $C$ (a so called partial
stem) and this partial theory should be obtained from the full one by
restriction.  Conversely, if the theory is formulated initially on the
stems, then the need arises for consistency between the partial theories
belonging to the 
different
stems of $C$.  
(From a histories point of view, almost all quantum theories are built
up this way from partial theories which provide the measures of
so-called cylinder events.  See for example [16] or
[17].)  
It is therefore an important question whether the theory herein
can be, or can be adapted to be, a component of a coherent theory
defined consistently on an untruncated causal set.

The other general question concerns interactions.  The theory we have
been working with herein is that of the causet counterpart of a free
field, as 
manifested
in the quadratic nature of (19).  In
its original operator form, however, the theory does not readily suggest
a generalization to include interactions.  A potential advantage of the
histories formulation, then, is that it does suggest such a
generalization.  To incorporate a $\phi^4$ interaction, for example, one
need only include a multiple of $i\xi^4-i\xibar^4$ into the exponent of
(19).  For the sake of consistency, one should check that the
decoherence functional, thus modified, remains positive semidefinite,
but this can be done.  It would be interesting to compare the
consequences of this generalized theory with those of its continuum
counterpart.


\section{Appendix.~ Proof of an identity used in the text}    
In proving equation (8) we used the following identity
(20), which is a special case of Wick's theorem.  Indeed, the
coefficient
$(n-1)!!=(n-1)(n-3)\cdots 5 \cdot 3 \cdot 1$ 
in (20) is easily seen to count the
number of ways of pairing up the $2n$ factors of $\Phi$ in the expression
$\bra \Phi^n \ket = \bra \Phi \Phi \Phi \cdots \Phi \ket$.   
Here we prove this special case directly by a straightforward method.
$$
    \mean{\Phi^{2n}} = (2n-1)!!  \; \mean{\Phi\Phi}^n   \eqno(20)
$$
Suppose first that 
$\Phi=a+a^*$ with $a$ normalized as usual so that $[a,a^*]=1$.  
Also let $|0\ket$ be the ``vacuum'' relative to $a$ with 
$\bra \cdot \ket = \bra 0| \cdot |0\ket$.
Noticing that $a|0\ket=0$ and $\bra0|a^*=0$, 
and that $[a,\Phi]=[a,a+a^*]=[a,a^*]=1$,
we have then
$$ 
 \bra \Phi^n \ket 
  = 
 \bra (a+a^*)\Phi^{n-1} \ket
  = 
 \bra a \Phi^{n-1} \ket
  = 
 \bra [a , \Phi^{n-1}] \ket
  = 
 \bra (n-1)\Phi^{n-2}\ket
  = 
  (n-1) \bra\Phi^{n-2}\ket  \ ,
$$
where we used the fact that $[a,\cdot]$ is a derivation (it obeys the
Leibniz rule). 
Since  $\bra \Phi^0 \ket =1$, it follows by induction that 
for even $n$, $ \bra \Phi^n \ket = (n-1)!!$.  When $n$ is odd 
$ \bra \Phi^n \ket$ of course vanishes, since an odd number of
applications of $a$ or $a^*$ to the vacuum can never bring one back to
the vacuum. 

Finally, consider the more general situation where $\Phi$ is some linear
combination of field operators $\phi(x)$~.
Because $\Phi=\Phi^*$ we can always write it in the form
$\phi=A+A^*$, where $A$ contains only the terms with lowering operators
$a$.  The proof then proceeds as before except that in place of 
$[a,a^*]=1$ we have instead 
$[A,A^*]=\mean{\Phi\Phi}$, as we see from
$\mean{\Phi\Phi}=\mean{(A+A^*)(A+A^*)}=\mean{A A^*}=\mean{[A,A^*]}=[A,A^*]$~.


\vskip 1 truecm
\bigskip
\noindent
For very helpful discussion of, and feedback on, the histories
formulation presented herein, I would like to thank Niayesh Afshordi and
in addition the participants of the meeting on causal sets held
December, 2009 at the Dublin Institute of Advanced Studies.
Research at Perimeter Institute for Theoretical Physics is supported in
part by the Government of Canada through NSERC and by the Province of
Ontario through MRI.

\ReferencesBegin

\ref [1] Luca Bombelli, Joohan Lee, David Meyer and Rafael D.~Sorkin, ``Spacetime as a Causal Set'', 
  \journaldata {Phys. Rev. Lett.}{59}{521-524}{1987}

\ref [2] Rafael D.~Sorkin, ``Causal Sets: Discrete Gravity (Notes for the Valdivia Summer School)'',
in {\it Lectures on Quantum Gravity}
(Series of the Centro De Estudios Cient{\'\i}ficos),
proceedings of the Valdivia Summer School, 
held January 2002 in Valdivia, Chile, 
edited by Andr{\'e}s Gomberoff and Don Marolf 
(Plenum, 2005)
\lbr
\eprint{gr-qc/0309009}
\lbr
\eprint{http://www.pitp.ca/personal/rsorkin/some.papers/} 

\ref [3] Joe Henson, ``The causal set approach to quantum gravity''
 \eprint{gr-qc/0601121}
 This is an extended version of a review published in 
 in  {\it Approaches to Quantum Gravity -- Towards a new understanding of space and time}, 
 edited by Daniele Oriti 
 (Cambridge University Press 2009)
 (ISBN: 978-0-521-86045-1), pages 26-43,

\ref [4] Fay Dowker, ``Causal sets and the deep structure of Spacetime'', 
 in
 {\it 100 Years of Relativity - Space-time Structure: Einstein and Beyond}"
 ed Abhay Ashtekar 
 (World Scientific 2005)
 \eprint{gr-qc/0508109}

\ref [5] Petros Wallden, ``Causal Sets: Quantum gravity from a fundamentally discrete spacetime'',
 \journaldata {J. Phys. Conf. Ser.}{222}{012053}{2010}
 \eprint{arXiv:1001.4041v1 [gr-qc]}

\ref [6] Sumati Surya,  ``Directions in Causal Set Quantum Gravity'',
 \eprint{arXiv:1103.6272v1 [gr-qc]}

\ref [7] Rafael D.~Sorkin, ``Logic is to the quantum as geometry is to gravity''
 in G.F.R. Ellis, J. Murugan and A. Weltman (eds),
 {\it Foundations of Space and Time} 
 (Cambridge University Press, to appear)
 \arxiv{arXiv:1004.1226 [quant-ph]}
 \eprint{http://www.pitp.ca/personal/rsorkin/some.papers/} 
 \sepref
Rafael D. Sorkin, ``An exercise in `anhomomorphic logic'~'',
 \journaldata{Journal of Physics: Conference Series (JPCS)}{67}{012018}{2007},
 a special volume edited by L. Diosi, H-T Elze, and G. Vitiello, and
 devoted to the Proceedings of the DICE-2006 meeting,
  held September 2006, in Piombino, Italia.
  \eprint{arxiv quant-ph/0703276} ,
 \lbr
 \eprint{http://www.pitp.ca/personal/rsorkin/some.papers/}

\ref [8] Rafael D. Sorkin, ``Does Locality Fail at Intermediate Length-Scales?''
 in  {\it Approaches to Quantum Gravity -- Towards a new understanding of space and time}, 
 edited by Daniele Oriti 
 (Cambridge University Press 2009)
 (ISBN: 978-0-521-86045-1), pages 26-43,
 \eprint{gr-qc/0703099},
 \eprint{http://www.pitp.ca/personal/rsorkin/some.papers/}

\ref [9] Don Marolf, private communication

\ref [10] Steven Johnston, ``Feynman Propagator for a Free Scalar Field on a Causal Set''
 \journaldata{Phys. Rev. Lett}{103}{180401}{2009}
 \eprint{arXiv:0909.0944 [hep-th]}

\ref [11] 
 Xavier Martin, Denjoe O'Connor and Rafael D.~Sorkin,
``The Random Walk in Generalized Quantum Theory''
\journaldata {Phys. Rev.~D} {71} {024029} {2005}
\eprint{gr-qc/0403085}
\lbr
\eprint{http://www.pitp.ca/personal/rsorkin/some.papers/}

\ref [12] Steven Johnston, ``Particle propagators on discrete spacetime''
\journaldata{Class.Quant.Grav.}{25}{202001}{2008}
\arxiv{0806.3083 [gr-qc]}

\ref [13] Dionigi M.T. Benincasa and Fay Dowker, ``The Scalar Curvature of a Causal Set''
\eprint{arXiv:1001.2725}
\journaldata{Phys.Rev.Lett.}{104}{181301}{2010}

\ref [14] Rafael D. Sorkin, ``Quantum dynamics without the wave function''
 \journaldata{J. Phys. A: Math. Theor.}{40}{3207-3221}{2007}
  (http://stacks.iop.org/1751-8121/40/3207)
\eprint{quant-ph/0610204} 
\lbr
\eprint{http://www.pitp.ca/personal/rsorkin/some.papers/}

\ref [15] Rafael D.~Sorkin, ``Quantum Mechanics as Quantum Measure Theory'',
   \journaldata{Mod. Phys. Lett.~A}{9 {\rm (No.~33)}}{3119-3127}{1994}
   \eprint{gr-qc/9401003}
   \lbr
   \eprint{http://www.pitp.ca/personal/rsorkin/some.papers/}


\ref [16] S.~Gudder and Rafael D.~Sorkin, ``Two-site quantum random walk'',
 {\it General Relativity and Gravitation} (to appear)
 \arxiv{1105.0705}
 \lbr
 \eprint{http://www.pitp.ca/personal/rsorkin/some.papers/}

\ref [17] Rafael D.~Sorkin, ``Toward a `fundamental theorem of quantal measure theory'$\,$'',
 {\it Mathematical Structures in Computer Science}
 (to appear)
 \lbr
 \eprint{http://arxiv.org/abs/1104.0997}
 \lbr
 \eprint{http://www.pitp.ca/personal/rsorkin/some.papers/141.fthqmt.pdf} 

\end                                         


(prog1 'now-outlining
  (Outline* 
     "\f"                   1
      "
      "
      "
      "
      "\\Abstract"          1
      "\\section"           1
      "\\subsection"        2
      "\\appendix"          1       ; still needed?
      "\\ReferencesBegin"   1
      "
      "\\ref "              2
      "\\end